# Charge transport in disordered organic solids: A Monte Carlo simulation study on the effects of film morphology

S. Raj Mohan, M. P. Joshi[*] and Manoranjan P. Singh

Laser Physics Applications Division,

Raja Ramanna Centre for Advanced Technology, Indore, MP - 452 013, India

* email: *mukesh@cat.ernet.in*

## Abstract

The influence of ordered regions (micro crystallites and aggregates) in the other wise disordered polymer host matrix on field and temperature dependence of mobility ($\mu$) has been simulated. Increase in concentration of ordered regions leads to increase in magnitude of mobility and in high field regime the saturation of the mobility occurs at lower electric field strength. The influence of different mean and standard deviation of Gaussian density of states (DOS) of ordered regions on the field dependence of mobility was studied and found to be significant only at higher concentrations. Weak influence of these parameters at low concentrations are attributed to the strong interface effects due to the difference in the standard deviation of DOS of two regions (host and ordered region) and shallow trapping effect by ordered regions. For all the parameters of ordered regions under investigation the temperature dependence of mobility ($\log\mu$) and the slope of $\log\mu$ Vs $E^{1/2}$ plot show $1/T^2$ dependence.

# Introduction

Thin films of conjugated organic solids (both molecular and polymers) have gained tremendous importance due to their potential applications in developing various optoelectronic devices like organic light emitting diodes (OLED), organic field effect transistors (OFET), organic solar cells etc[1,2]. A better understanding and control of morphology of disordered organic solid films is of prime importance because the performance of several organic devices have been shown to be highly sensitive to film morphology and processing conditions[3-8]. Often, the active layers of devices, e.g. spin cast films of molecularly doped polymers (MDP) and conjugated polymers, are not purely disordered or amorphous rather they are partially ordered. Ordered regions (e.g. molecular aggregation and crystallization in MDP or oriented polymer chains in conjugated polymers) are formed either unintentionally due to processing conditions and aging or intentionally incorporated via annealing. Compared to purely amorphous films the partially ordered polymer films show an improvement in carrier mobility (μ) and charge transport properties. Therefore a careful control of deposition parameters and processing conditions [3, 9-12] are generally employed to improve the structural order in these materials. Charge transport in these films occur through a mixture of ordered and less ordered regions, which is against the assumptions made in hopping charge transport models developed for completely isotropic and disordered medium [13-15]. In these models of charge transport the influence of film morphology of the active layers, which has profound effects on charge transport, is not considered well. The presence of ordered regions can in-fact reduce the overall energetic disorder in the material [9,16] that can enhance the mobility and at the same time can substantially influence the behavior of charge transport[17-20]. The presence of structural order can change the mechanism of charge transport drastically because it may be possible that in ordered regions the charge transport may be of the band type and hence the overall mechanism may be a combination of the band and the hopping transport [21]. Field dependence of mobility in partially ordered samples generally shows a Poole-Frenkel type behavior [22-25] but in some cases either very weak field dependence or even negative field dependence, at low temperatures, of mobility have been observed [16,18,22,26,27]. Temperature dependence of mobility in these partially ordered films has also been a matter of discussion whether logμ follows $1/T$ or $1/T^2$. Experiments as well as simulations report $1/T$ [26-28] as well as $1/T^2$ dependence [23] and some cases $1/T$ and $1/T^2$ dependence of mobility (logμ) are indistinguishable [24]. In general, a proper understanding of charge transport in these partially ordered materials is yet to be obtained which is important for developing appropriate model for explaining the charge transport as well as for controlling the morphology of active layers.

In this study we perform Monte Carlo simulation to investigate the influence of ordered regions on the charge transport when incorporated in a disordered host lattice. The problem is similar to charge transport in samples like molecularly doped polymers (MDP) where sub micron sized microcrystallization/aggregation of dopants occur unintentionally or intentionally [18,29,30] or when partial structural ordering (like alignment of polymer chains) takes place in amorphous polymer films [4,6,10]. Our observation of aggregates of *N-N'-diphenyl-N-N'-bis(3-methylphenyl)-1-1- biphenyl-4,4'-diamine* (TPD) in TPD doped polymers and the significant influence of these aggregates (of size ~100nm) on charge transport [18] also motivated us to simulate the charge transport mechanism in partially ordered films and provide justification for our experimental results. In the simulation, the morphology of the disordered host lattice is altered by randomly incorporating submicron sized cuboids with a narrow energetic distribution of localized states (ordered region). Simulation is performed for various electric field strength and temperature by changing the concentration or ordered regions, standard deviation and mean energy of density of states (DOS) of ordered regions. The influence of these parameters is important because the variation in these parameters is common as the degree of order obtained in the active layers vary depending on the processing conditions.

The incorporation of ordered regions leads to increase in magnitude of mobility with increase of concentration of ordered regions while at high field regime the saturation of mobility occurs at lower electric field strength. The influence of parameters like standard deviation and mean energy of DOS of ordered regions is found to be significant only at high concentration. While at low concentrations these parameters have weak influence because of interface trap effects which arise due to difference in standard deviation of two regions. The temperature dependence of mobility and the slope of log$\mu$ Vs. $E^{1/2}$ plot at intermediate field regime follow $1/T^2$ as predicted by Gaussian disorder model (GDM) [14].

**Simulation procedure**

The Monte Carlo simulation is based on the commonly used algorithm reported by Schönherr et al [31]. A lattice of 70x70x300, along x, y and z direction, with lattice constant *a* =

6Å was used for computation. Z direction is considered as the direction of the applied field. The size of the lattice is judged on the basis of our aim to change the morphology of the sample and also by taking into account the available computational resources. The site energies of lattice were taken randomly from a Gaussian distribution of mean ~5.1eV and standard deviation σ = 75meV, which gives the energetic disorder parameter $\hat{\sigma} = \sigma / kT$. The value for σ and mean were chosen close to the experimental values observed in TPD based MDPs. Simulation was performed on this energetically disordered lattice with the assumption that the hopping among the lattice sites are controlled by Miller-Abrahams equation [32] in which the jump rate $\upsilon_{ij}$ from the site $i$ to site $j$ is given by

$$\nu_{ij} = \nu_0 \exp\left(-2\gamma a \frac{\Delta R_{ij}}{a}\right) \begin{cases} \exp\left(-\frac{\varepsilon_i - \varepsilon_j}{kT}\right) \exp\left(\pm\frac{eEa}{kT}\right) &, \varepsilon_j > \varepsilon_i \\ 1 &, \varepsilon_j < \varepsilon_i \end{cases} \quad (1)$$

where $E$ is the applied electric field, $a$ is the intersite distance, $k$ is the Boltzmann constant, $T$ is the temperature in Kelvin, $\Delta R_{ij} = |R_i - R_j|$ is the distance between sites $i$ and $j$ and $2\gamma a$ is the wave function overlap parameter which controls the electronic exchange interaction between sites. Throughout the simulation we assume $2\gamma a = 10$ [14,31].

With the aim to determine the transit time , the time taken by a carrier to cover the entire sample length defined in the simulation, a carrier residing randomly in the first plane (z=0) was allowed to move in the lattice under the action of applied field. Hopping of carrier from site $i$ to site $j$ was performed on the basis of the probability that a carrier jumps from the present site $i$ to any site $j$ around and within a cube of size 7x7x7 (343 sites). Probability of jump $P_{ij}$ is given by,

$$P_{ij} = \frac{\nu_{ij}}{\sum_{i \neq j} \nu_{ij}} \quad (2)$$

A random number $u_r$ from a uniform distribution is chosen and this decides to which site the carrier should jump because each site is given a length in random number space according to $P_{ij}$. The time taken by a carrier for jumping from site $i$ to $j$ is given by,

$$t_{ij} = \frac{x_{\exp}}{\sum_{i \neq j} v_{ij}} \qquad (3)$$

where $x_{\exp}$ is a random number taken from exponential distribution. Using periodic boundary condition along x, y and z direction the simulation was always performed for a sample of length L ~ 4μm along the field direction. The sample length was chosen so as to make sure that carrier will attain thermal equilibrium during its transit [14]. Whenever periodic boundary condition was used the lattice configuration was changed to ensure that the carriers do not move again through the same lattice. In the present study we calculated mobility ($\mu = L^2/V\tau$) from the average transit time ($\tau$) obtained after averaging over 150 carriers with one lattice configuration for each carrier. Simulation was preformed for various electric field and temperature.

Film morphology was varied by incorporating cuboids of so called ordered regions (representing molecular aggregates/microcrystallites in MDPs or regions of oriented polymer chains in conjugated polymers) of varying size that are placed randomly inside the otherwise disordered host lattice. Sizes of ordered regions were limited to a maximum size of 12x12x100 nm along x, y and z directions. The maximum size was chosen so as to simulate the charge transport in a situation that is close to experimentally observed TPD aggregates [18]. The energetic disorder inside the ordered region was kept low compared to the host lattice. This is justified because the aggregates/microcrystallites are more ordered regions and hence to simulate the charge transport the cuboids must be of low energetic disorder compared to host lattice. The site energies inside the ordered regions were also taken randomly from another Gaussian distribution. Aggregation of dopants can also lead to change in the energy gap (shift in HOMO, LUMO levels) and can change the mean energy of Gaussian distribution. The mean energies of ordered regions were chosen such that their difference from the mean energy of host lattice is in the order of kT. Simulations were performed by varying the concentration of such ordered regions (varying the percentage of volume of lattice occupied by ordered region) and the standard deviation and mean of Gaussian distribution from which the site energies of ordered regions were assigned.

**Results and Discussions**

Fig. 1 show the field dependence of mobility simulated for a lattice having DOS with standard deviation ~75meV (the host lattice), parametric with various values of energetic disorder parameter ($\hat{\sigma}$). At low field strength the mobility shows saturation while at intermediate field strength the mobility increases with increase of electric field in $\log\mu$ Vs. $E^{1/2}$ fashion. This is because the applied field tilts the density of states that lead to the decrease of energetic barrier which the carriers encounter [14]. When temperature is increased carriers gain thermal energy and hence the effect of energetic disorder decreases which results in decrease of the slope of $\log\mu$ Vs. $E^{1/2}$ curve at intermediate field strengths. At higher electric field strength the drift velocity of charge carrier saturates and mobility attain a maximum value and decreases with further increase of electric field in a 1/E fashion. At higher temperature the drift velocity saturates at low electric field strength leading to the saturation of mobility [14]. Inset of the Fig. 1 shows the remarkable difference in the field dependence of mobility for a lattice having DOS with standard deviation 60mev, 40meV and 2meV. With decrease in energetic disorder it is observed that mobility increases and the slope of mobility curve, at intermediate field strength, changes from positive to negative.

To study the influence of embedded micro crystals/aggregation of dopants the simulation was performed after incorporating energetically ordered regions inside the host lattice (as explained in simulation procedure). Fig. 2 shows the field dependence of mobility, at ~248K, parametric with the concentration of ordered regions having DOS with standard deviation ~40meV and mean energy lower by ~ kT compared to mean energy of host lattice. Magnitude of mobility at all regimes of electric field except the high field regime increases with increase in the concentration of ordered regions concomitant with decrease of slope at the intermediate field regime. In high field regime, the incorporation of ordered regions leads to the saturation of the mobility at lower electric field strength. When the concentration of ordered region is increased, the saturation of mobility also occurs at lower electric field strength. These observed features in the field dependence of the mobility, after embedding the ordered regions in the host lattice, can be rationalized on the basis of decrease in effective energetic disorder of the transport medium. At low value of effective energetic disorder the energetic barrier seen by the carriers will be less which results in higher mobility but a weaker field dependence. This explains the increase in magnitude of mobility and decrease in the slope of logμ Vs. $E^{1/2}$ curve at intermediate field regime with increase in concentration of ordered regions in the host lattice. The low value of energetic disorder also can lead to saturation of drift velocity and mobility at lower electric field strength.

From Fig. 2 it is clear that the magnitude of mobility attained even for concentration of ordered region as high as 80% is still ~ 8 to 10 times less than that of a host lattice with DOS same as that of ordered region (but without any crystallites). Inset of Fig. 2 shows the variation of mobility (at ~248K and field ~1.44 x $10^6$ V/cm) with concentration of ordered region, having DOS with standard deviation ~ 40meV. A remarkable nonlinear increase in the mobility was observed only when the concentration of the ordered regions was greater than ~60%. According to percolation theory this is expected to occur at ~25-30% concentration [33]. The observation of large threshold and weakly percolating transport is due to the fact that when a group of carriers moves in a mixture of high and low mobility regions then only few carriers may find very low resistance paths (paths containing very low fraction of low mobility regions) with short transit time and overall high mobility only for those carriers. To obtain further insight into this aspect the path of the 250 carriers, for various concentration of ordered region inside the host lattice, was followed and the fraction of path occupied by the ordered region was calculated. Simulation was performed with one configuration for each carrier. From the histogram, shown in Fig. 3 for various concentrations of ordered region inside the host lattice, it is clear that very few low resistant paths (paths with fraction of ordered regions >0.95) of high mobility exist in simulated lattice. The number of such low resistance paths is very less compared to those paths where the fraction of low mobility region is considerable, i.e. high resistant (low mobility) paths. Hence on averaging over number of carriers the mobility value will be dominated by contribution from those high resistant (low mobility) paths which results in low magnitude of average mobility compared to uniform lattice with DOS of standard deviation same as that of ordered region. Moreover, the interface effect which occurs at the interface between ordered region and host lattice due to the difference in standard deviation of DOS also increases the transit time. Interface effects will be discussed later in this paper. Increase in fraction of less resistant paths with increase of concentration of ordered region in the host lattice also supports the improvement in the mobility with increase in concentration of ordered region in host lattice.

The standard deviation and mean energy of DOS for ordered regions also influence the charge transport, especially the field and temperature dependence. The influence of disorder inside the ordered region on charge transport properties is investigated by performing simulation by varying the standard deviation of DOS of the ordered region and the concentration of such ordered regions. Fig. 4 shows comparison of the field dependence of mobility for various values of the concentration and the standard deviation of DOS of ordered regions. Higher mobility is expected when ordered regions have a low value energetic disorder. It was, however, observed that the field dependence of mobility did not vary significantly with the change in the energetic disorder of the ordered region as long the concentration was below 60%. As we show below this happens because of the presence of the interface between the ordered region and the host lattice which acts like a "trap'' for a carrier. Once carrier enter the ordered region, where the energetic disorder is low and mobility high, it spends less time to traverse the entire length of ordered region along the field direction. However at other interface the carrier face energetic barrier and may undergo several back and forth jumps before it proceeds forward through the host lattice. Thus, at the second interface carrier looses some part of the time that it has gained while traversing inside the ordered region. The interface effect is more prominent for ordered regions with very low energetic disorder. Hence on average the total transit time of a carrier (the time taken to traverse the entire lattice), effectively the carrier mobility, for different values of energetic disorder inside the ordered region may not vary appreciably as long as concentration is low. To elucidate this interface effect a single ordered region was sandwiched between regions of high energetic disorder (as shown in the inset of Fig. 5). Region 2 is ordered region with reduced $\sigma$ while regions 1 and 3 are the regions with fixed $\sigma = 75$meV (all regions are of same dimension, 70x70x150). Carrier transit time is recorded from the first plane of the first region. From the time carrier first enters and first comes out from the region 2, the transit time of the carrier inside the region 2 (ordered region) was calculated. Similarly from the time carrier first comes out of region 2 and the time it reached the other end of the lattice, the transit time of the carrier in region 3 was also calculated. Simulation was performed by varying disorder inside the region 2. The time was calculated after averaging over 2000 carriers with one configuration for each carrier. Fig. 5 shows typical values of carrier transit time in region 2 and 3 for various values of energetic disorder in region 2. The dependence of carrier transit time in region 3 on the energetic disorder in region 2 was clearly observed. Carrier transit time in region 3 increases with decrease of energetic disorder inside the region 2. This observation can be completely attributed to the interface effects. When the disorder inside the region 2 decreases then barrier seen by the carrier at the interface of region 2 and 3 increases. Basically at the interface the carrier will hop back and forth several times before moving forward through the region 3. This results in the increase of carrier transit time in the region 3. Transit time of carrier inside region 2 increases with increase of energetic disorder inside the region 2 as expected from GDM. This observation supports our explanation of almost independent nature of field dependence of mobility on concentration of ordered regions, upto 60 % of ordered regions (as shown in Fig. 4), on the basis of interface effects. At high concentration, due to close vicinity of ordered regions, carrier spends most of the time in the ordered regions with less interface effects. Therefore mobility will be higher when the ordered region has lower energetic disorder and vice versa. Thus at higher concentrations the charge transport is mainly governed by the energetic disorder inside the ordered region. This behavior was shown in our simulation for high concentrations of ordered regions (shown in Fig. 3 for 80% of ordered region in our study). We have also observed that with ordered region incorporated in host lattice the slope of log$\mu$ Vs $E^{1/2}$ plot at intermediate field regime changes from positive to negative at lower temperatures (figure not shown) compared to purely disordered lattice. This was attributed to the overall low value of energetic disorder due to the presence of ordered region in the host lattice.

Simulations were also performed with different mean energy for DOS of ordered region compared to the host lattice. As explained in earlier section that the mean energy may change (normally decrease) upon aggregation, crystallization or molecular ordering. Also, there are examples of molecular species of low ionization potential dispersed in hole transporting host material. It has been shown that in such systems the dopant act as trap at low concentrations while at high concentrations the entire charge transport occurs through hopping among dopant species [34]. We performed simulation for three cases; (1) the mean energy of ordered region less than that of host lattice, (2) the mean energy of ordered region same as that of host lattice and (3) the mean energy of ordered region higher than that of host lattice. Simulation was performed for a given concentration of ordered region and varying the mean energy in the order of kT. We chose three different concentrations of ordered regions namely 20, 40 and 80%. As shown in Fig. 6, at low field and low concentration, mobility is little higher for case (1) than for case (2) and smaller for case (3). With increasing concentration of ordered regions this difference in mobility is seen more prominently. While at high field regime the field dependence of mobility in all the cases are nearly same. The observed differences in mobility for three cases are due to the fact that at all concentrations the low mean energy regions act as shallow traps. The low mean energy regions are either high mobility ordered region or low mobility host regions. For example, in case (1) the low mean energy regions are high mobility ordered regions. Shallow trapping in ordered region also helps the carrier in finding more suitable paths inside the ordered region so as to move fast along the field direction. While in case 3 the low mean energy region is low mobility host lattice. At very high concentrations (e.g. 80%) carrier will spend good amount of its transit time mostly inside the ordered region therefore the effects of shallow traps are seen more clearly as large difference in mobility behavior. In case (3) even at such high concentration the trapping effect due to the host lattice is still effective and responsible for low mobility. This trapping mechanism due to difference in mean energy of DOS is an additional trapping mechanism which coexists at interface due to difference in standard deviation of DOS. Thus the mobility in case (3) is low compared to case (1) and (2). This mechanism is further get clarified from the histogram shown in inset of the respective Fig. (6 (a-c)). This histogram shows the fraction of the ordered region in the different paths followed by carriers, for case (1) and case (3), at the respective concentrations. It is clear from histogram that carriers find more number of least resistant paths in case (1) than in case (3) which results in higher mobility for case (1).

Temperature dependence of mobility shown by such partially ordered active layers may be different from the predictions of GDM because of the fact that charge transport now occurs through ordered and disordered regions. According to the simulation work of Rakhmanova *et al* [28] who also carried out simulation on a energetically inhomogeneous lattice predicted that the zero field mobility follow $1/T$ temperature dependence. The lattice chosen in [28] was completely an inhomogeneous lattice where site energy at *each jump site* was chosen randomly from two Gaussians of different standard deviation. In contrast to prediction of [28] our simulations predict $1/T^2$ dependence for mobility. Fig. 7(a) shows temperature dependence of zero field mobility and slope of logµ Vs. $E^{1/2}$ curve at intermediate field regime for a host lattice with σ = 75 meV. The observed $1/T^2$ dependence is as predicted by the GDM. Even when ordered regions of different concentrations and different σ are included in our simulated lattice the temperature dependence of mobility are better fit with $1/T^2$ than $1/T$ (as shown in Fig. 7(b) and (c)). The absence of $1/T$ dependence of mobility in our simulation compared to earlier report [28] is because of different degree of inhomogenity in the two lattices. In the present study the ordered regions have sub micron sized spatial extensions and inhomogenity is seen only at interfaces. When carrier is either inside the ordered region or in the host lattice it moves in a region with site energies decided by a single Gaussian and transport in both the region should follow GDM. Due to large spatial extension of ordered regions the energetic inhomogeny seen by the carrier must be less compared to the simulated lattice used in earlier report [28]. Only at the interface of two regions carrier see energetic inhomogeny. Hence the absence of 1/T dependence of mobility in our simulation can be attributed to low amount of inhomogenity in the lattice. Our simulation suggests that the charge transport in the investigated lattice occurs in similar manner as predicted by GDM.

## Conclusion

We studied the influence of embedded ordered regions on the charge transport in otherwise disordered amorphous systems. This is similar to the case of aggregation of dopants in molecularly doped polymers or when the microcrystals are deliberately embedded inside the conjugated polymer matrix. The effect of embedded ordered regions is seen as decrease of overall energetic disorder of the system. This decrease in energetic disorder results in the increase of over all mobility, the decrease of slope of log$\mu$ Vs. $E^{1/2}$ curve at intermediate field regime and, at high field regime, the saturation of mobility occurs at lower field strength. The influence of standard deviation and mean energy of ordered region on the field dependence of mobility was seen as shallow trapping effect which occurs at the interface between ordered region and host lattice. At low concentration, due to shallow trapping effect, the field dependence of mobility do not show any significant change with standard deviation and mean energy of ordered regions. A remarkable influence of these parameters was observed only at high concentrations where mobility was higher when ordered regions have low energetic disorder and low mean energy compared to the host lattice. The temperature dependence of mobility and the slope of log$\mu$ Vs. $E^{1/2}$ curve follow $1/T^2$ dependence than 1/T dependence as predicted in earlier simulations for inhomogeneous lattice. This is attributed to low inhomogenity seen by the carrier in the simulated lattice. Simulation suggests that one can have higher mobility with less inhomogenity for carrier transport by having ordered regions inside the lattice.

## Acknowledgements

Authors are grateful to S. C. Mehendale for the support and encouragement. S. Raj Mohan is grateful to Department of Atomic Energy (DAE), India, for providing Senior Research Fellowship. SRM like to thank J. Jayabalan, Aparana Chakravarthi and V. K. Dixit of RRCAT for fruitful discussions.

# Figure Captions

Fig. 1. Field dependence of mobility for host lattice with $\sigma = 75$meV parametric with energetic disorder parameter $\hat{\sigma}$ = 3.5 ($\diamond$), 3 ($\blacktriangledown$), 2 ($\triangle$), 2.5 ($\bullet$), 1.5 ($\square$). Inset shows the field dependence of mobility for a lattice with $\sigma$ = 60meV ($\diamond$), 40meV ($\blacksquare$) and 2meV ($\blacktriangle$).

Fig. 2. Field dependence of mobility for a lattice, at 248K, with 0% ($\blacksquare$), 20% ($\bullet$), 40% ($\blacktriangle$), 60% ($\blacktriangledown$), 80% ($\blacklozenge$) and 100% ($\star$) concentrations of ordered regions having DOS with standard deviation ~40meV. Inset shows the variation of mobility with the concentration of ordered regions having DOS with standard deviation ~40meV, at 248K and at field E =1.44x10$^6$ (V/cm).

Fig. 3. Histograms showing the fraction of ordered regions in the independent path followed by 250 carriers for different concentration of ordered region inside the host lattice, at 248K and field E =1.44x10$^6$ (V/cm). Standard deviation of the DOS of ordered regions were taken to be ~40meV.

Fig. 4. Field dependence of mobility, at 248K, of a host lattice embedded with different concentrations of ordered regions having DOS with standard deviation ~60meV ($\blacksquare$), 40meV ($\circ$) and 2meV ($\blacktriangle$). For comparison the field dependence of host lattice without embedded ordered regions is also shown ($\star$).

Fig. 5. Dependence of transit time of carrier in the region 3 ($\bullet$) and region 2 ($\square$) as a function of $\sigma$ in region 2, at 248K and field=1.44x10$^6$ (V/cm). Geometry of three regions is shown in the inset. For region 1 and 3, $\sigma$ = 75meV.

Fig. 6. Field dependence of mobility of a host lattice, at 248K, with (a) 20%, (b) 40%, and (c) 80% concentration of ordered regions having different mean energy compared to the mean energy of host lattice. Case 1 ($\blacksquare$), Case 2 ($\bullet$), and Case 3 ($\blacktriangle$). Mean

energy difference in each case is shown in the upper inset of Figure 6(a). Lower inset of each figure shows the histogram of fractions of ordered regions in the independent path followed by 250 carriers with respective concentration of ordered regions having $\sigma = 40$meV, at 248K and $E = 1.44 \times 10^6$V/cm.

Fig. 7. (a) Temperature dependence of zero field mobility for a host lattice having DOS with standard deviation 75meV. Straight lines are the linear fit to the data. (b) and (c) are temperature dependence for a host lattice with 40% of sites occupied by ordered regions having $\sigma = 40$meV and 2meV respectively. Inset of each figure shows the temperature dependence of slope ($\beta$) of log$\mu$ Vs. $E^{1/2}$ plot at intermediate field.

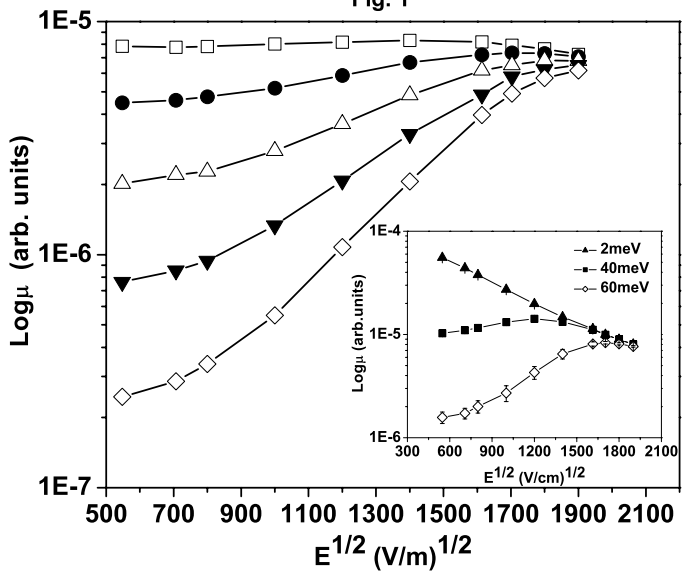

Fig. 1

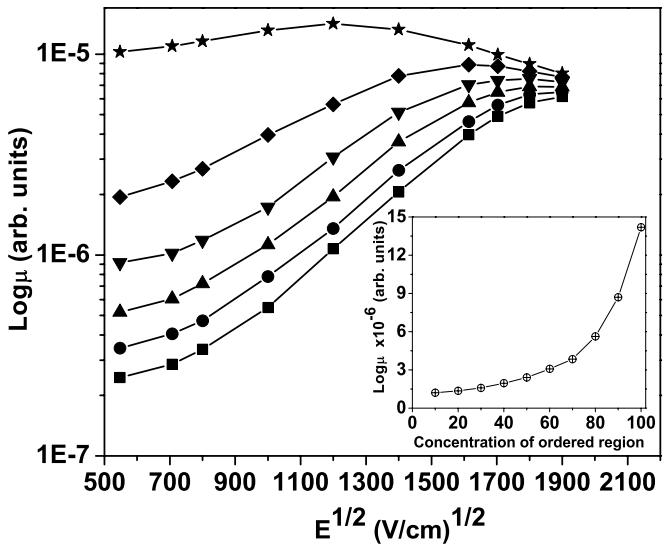

**Fig. 2**

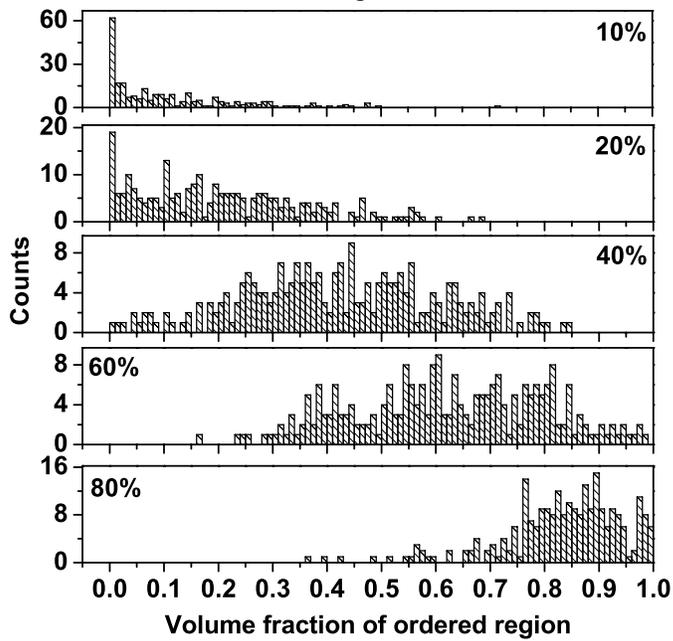

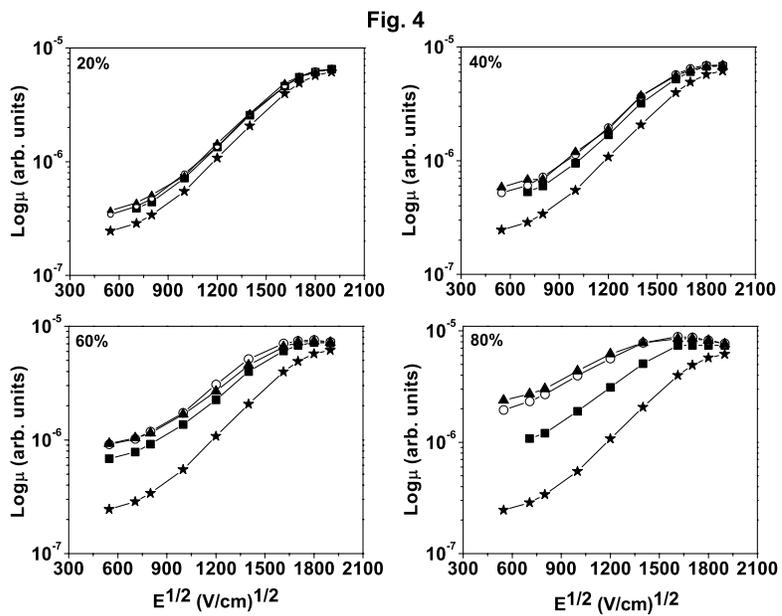

Fig. 4

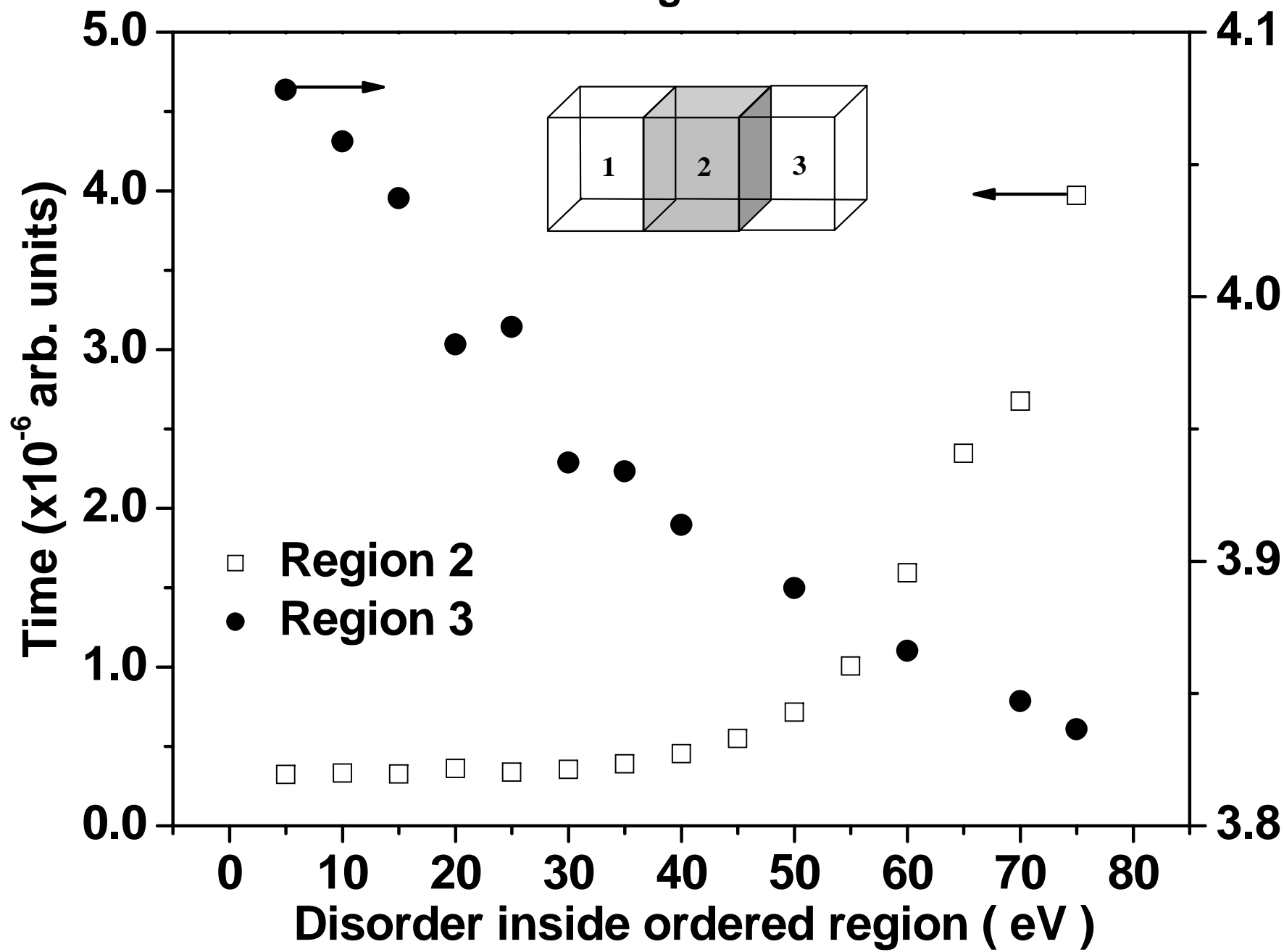
Fig. 5

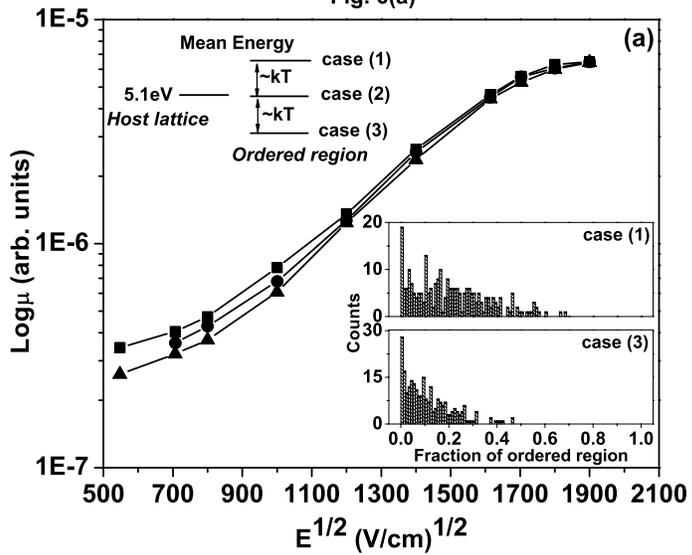

Fig. 6(a)

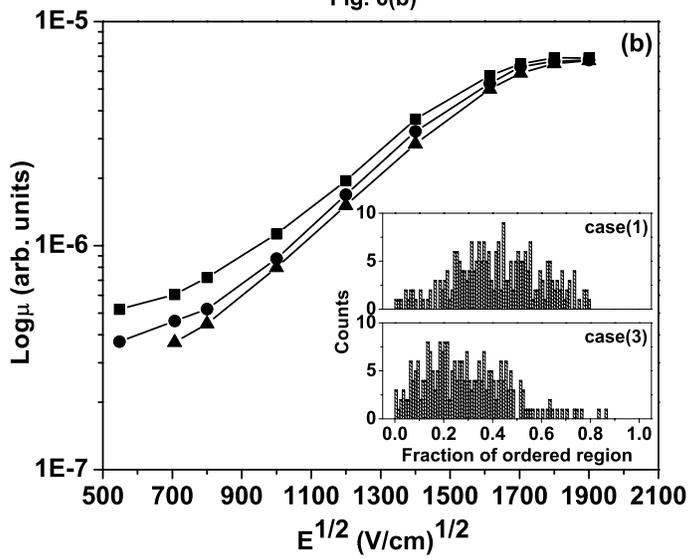

Fig. 6(b)

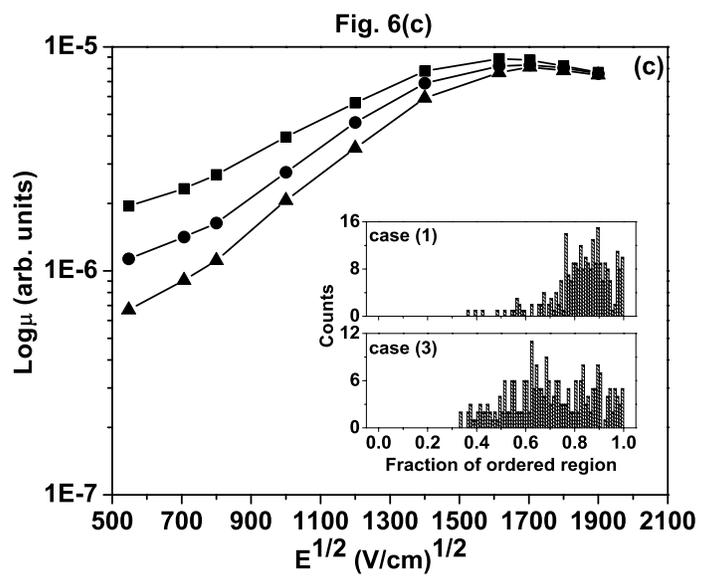

Fig. 6(c)

**Fig. 7(a)**

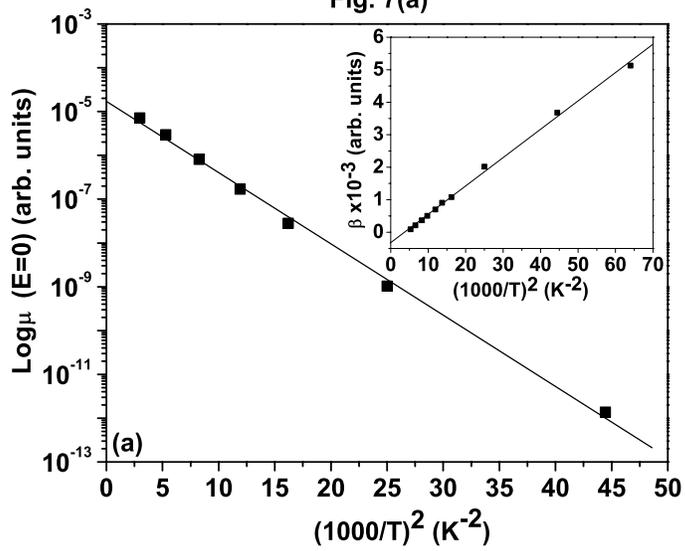

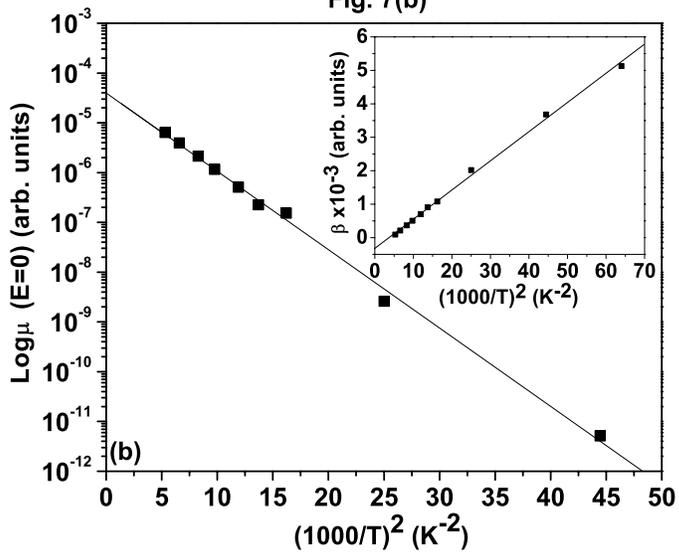

Fig. 7(b)

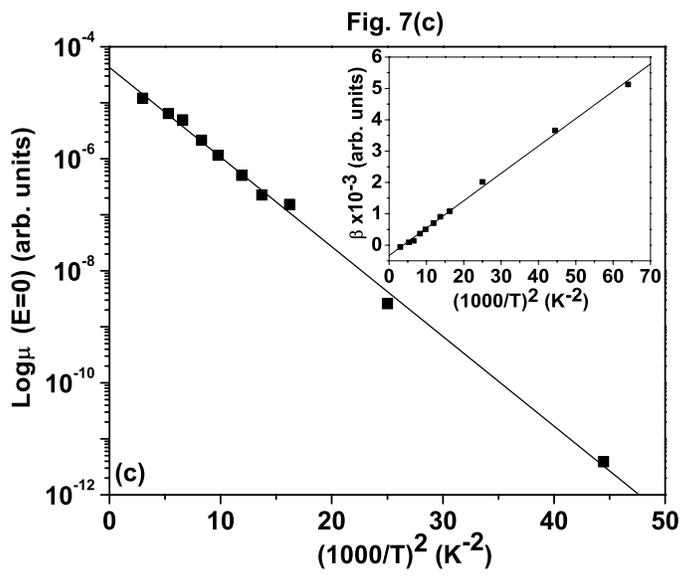

Fig. 7(c)